\documentclass{SciPost}
\usepackage{etoolbox}
\usepackage{mathtools}
\usepackage{dcolumn}% Align table columns on decimal point
\usepackage{bm}% bold math
\usepackage{verbatim}
\usepackage[english]{babel}
\usepackage{graphicx}%
\usepackage{multirow}%
\usepackage{amsmath}%
\usepackage[title]{appendix}%
\usepackage{xcolor}%
\usepackage{textcomp}%
\usepackage{booktabs}%
\usepackage{listings}%
\usepackage{bbold}
\usepackage{amsthm}
\usepackage[capitalise]{cleveref}
\usepackage{amsmath}
\hypersetup{
    colorlinks,
    linkcolor={red!50!black},
    citecolor={blue!50!black},
    urlcolor={blue!80!black}
}

\usepackage[bitstream-charter]{mathdesign}
\DeclareSymbolFont{usualmathcal}{OMS}{cmsy}{m}{n}
\DeclareSymbolFontAlphabet{\mathcal}{usualmathcal}

\fancypagestyle{SPstyle}{
\fancyhf{}
\lhead{\colorbox{scipostblue}{\bf \color{white} ~SciPost Physics }}
\rhead{{\bf \color{scipostdeepblue} ~Submission }}

\fancyfoot[C]{\textbf{\thepage}}
}

\definecolor{emerald}{rgb}{0.31, 0.78, 0.47}
\definecolor{blue(ncs)}{rgb}{0.0, 0.53, 0.74}

\begin{document}

\pagestyle{SPstyle}

\begin{center}{\Large \textbf{\color{scipostdeepblue}{
Resonating Valence Bond Ground States on Corner-sharing Simplices\\
}}}\end{center}

\begin{center}\textbf{
Zhao Zhang\textsuperscript{1$\star$} and
Cecilie Glittum\textsuperscript{1,2,3$\dagger$}
}\end{center}

\begin{center}

{\bf 1} Department of Physics, University of Oslo, P.O. Box 1048 Blindern, N-0316 Oslo, Norway
\\
{\bf 2} Helmholtz-Zentrum Berlin f\"ur Materialien und Energie GmbH, Hahn-Meitner-Platz 1, 14109 Berlin, Germany\\
{\bf 3} Dahlem Center for Complex Quantum Systems and Fachbereich Physik, Freie Universit\"at Berlin, 14195 Berlin, Germany
\\[\baselineskip]
$\star$ \href{mailto:email1}{\small zhao.zhang@fys.uio.no}\,,\quad
$\dagger$ \href{mailto:email2}{\small cecilie.glittum@fys.uio.no}
\end{center}

\section*{\color{scipostdeepblue}{Abstract}}
\textbf{\boldmath{
The Hubbard model in the $U\to\infty$ limit has been known to have resonating valence bond (RVB) ground states on certain corner-sharing simplex lattices. Examples include both the quasi-1D sawtooth lattice with open boundary and a larger class of higher dimensional lattices without boundaries. The two types of results were obtained by different approaches which do not apply to one another. In the second class of lattices, the simplest simplex is a tetrahedron. We hereby generalize both results by studying the singly hole-doped system on the quasi-1D lattice of a tetrahedron chain, which can be considered a stripe of the pyrochlore or checkerboard lattices. The energy level ordering of irreducible representations of each tetrahedron shows that a chain of them has exponentially degenerate partial RVB or dimer-monomer ground states where each tetrahedron hosts one spin-$1/2$ monomer and one spin-$0$ dimer. The exact ground states in the infinitely long chain limit are analytically solved by introducing basis transformations between local Hilbert spaces of neighboring tetrahedra, and its energy agrees with the extrapolation of numerical exact diagonalization results of finite sized systems.
}}

\vspace{\baselineskip}

%%%%%%%%%% BLOCK: Copyright information
% This block will be filled during the proof stage, and finilized just before publication.
% It exists here only as a placeholder, and should not be modified by authors.
\noindent\textcolor{white!90!black}{%
\fbox{\parbox{0.975\linewidth}{%
\textcolor{white!40!black}{\begin{tabular}{lr}%
  \begin{minipage}{0.6\textwidth}%
    {\small Copyright attribution to authors. \newline
    This work is a submission to SciPost Physics. \newline
    License information to appear upon publication. \newline
    Publication information to appear upon publication.}
  \end{minipage} & \begin{minipage}{0.4\textwidth}
    {\small Received Date \newline Accepted Date \newline Published Date}%
  \end{minipage}
\end{tabular}}
}}
}
%%%%%%%%%% BLOCK: Copyright information

% For convenience during refereeing we turn on line numbers:
%\linenumbers

\vspace{10pt}
\noindent\rule{\textwidth}{1pt}
\tableofcontents
\noindent\rule{\textwidth}{1pt}
\vspace{10pt}

\section{Introduction}
\label{sec:intro}

The resonating valence bond (RVB) state is a competing ground state to N\'eel antiferromagnetic (AFM) order. It was proposed by Anderson in 1973~\cite{ANDERSON1973153} in analogy to Pauling's attempted theory of metals abandoning the Fermi gas picture and motivated by Bethe's solution of the spin-$\frac{1}{2}$ AFM Heisenberg chain \cite{1931_Bethe_ZP_71}, which shows a liquid-like ground state without sublattice magnetization in the absence of anisotropy. 
The hypothesis of quantum spin liquidity \cite{balent} regained interest following the discovery of high-temperature superconductivity, as doping allows the system to metalize and the existing real-space `Cooper pairs' to Bose condense \cite{HighTcRVB}. 

Short-range RVB states have been modeled by close-packed quantum dimers on square \cite{PhysRevLett.61.2376} and triangular lattices \cite{PhysRevLett.86.1881} under ring-exchange interactions, showing drastically different excitation spectra.\footnote{It should be noted that the plaquette flipping move alone on the triangular lattice leads to Hilbert space fragmentation, resulting in exponentially large ground state degeneracy \cite{10.21468/SciPostPhysCore.6.3.054}. For bipartite lattices like the square lattice, due to the emergent Coulomb gauge, there is a well-defined height function on the dual lattice that leaves ground state degeneracy to be accounted for only by topological sectors \cite{Zhang_2023}. This could be the reason behind the difference between a gapped and a gapless spectrum.} By construction, the ground states of quantum dimer models become frustration-free RVB states at the Rokhsar-Kivelson (RK) point. However, the orthogonal dimer basis in the Hilbert space of dimer configurations is quite different from the overcomplete basis of spin singlet pairs formed by spin-$\frac{1}{2}$'s. While RK moves can also be combined with Klein Hamiltonians \cite{Klein_1982} that projects onto non-dimerized SU(2) states to realize frustration-free spin Hamiltonians \cite{PhysRevLett.105.067205}, the interaction has to involve a neighborhood of eight spins on a square lattice.

Perhaps a more physically realistic route towards an exact RVB state is via frustration of the kinetic energy of itinerant electrons~\cite{PhysRevLett.70.3303, pyrochlore}, as opposed to localized spin interactions. Itinerant electrons are known to induce effective spin interactions, for example in the form of Nagaoka ferromagnetism \cite{Thouless_1965, PhysRev.147.392}, appearing in infinite-$U$ Hubbard models on generic bipartite lattices at half-filling upon the doping of a single hole (or electron). If the expectation value of the hopping Hamiltonian is minimized by an antisymmetric spatial wavefunction between an electron and the holon, their spin configurations must be symmetric to satisfy fermionic statistics. Since the electron spins in the system are all antisymmetrized with the holon, their spins must be aligned with one other.

On a non-bipartite lattice, due to kinetic frustration, the spatial wavefunction of the electrons cannot be antisymmetrized. So a symmetric spatial wavefunction that minimizes the kinetic energy with negative hopping strength results instead in the AFM ordering of the spins \cite{PhysRevLett.95.087202, PhysRevLett.112.187204}, referred to as the counter-Nagaoka effect. The idea that infinite-$U$ Hubbard models on such frustrated lattices can induce exact antiferromagnetic ground states with localized dimers has been extensively explored in the 1990s, when Brandt and Giesekus established the result analytically for decorated hypercubic lattices in dimensions higher than two \cite{PhysRevLett.68.2648}. The paradigm was further generalized to non-uniform hopping \cite{PhysRevLett.70.3303}, and to quasi-1D tetrahedron chain with additional onsite potential \cite{PhysRevB.52.2476}. This series of results will be reviewed in Appendix \ref{sec:BG}.

The class of lattices considered by Brandt and Giesekus is actually a special case of corner-sharing simplices, which are line graphs of the lattices of the simplex centers. Mielke analyzed the graph theoretic properties of these lattices, and discovered that the frustration-free ground states of Brandt and Giesekus can be generalized to hole-doped systems \cite{Mielke_1992}. This is a major revelation as non-trivial results already apply to corner-sharing tetrahedron lattice, with realistic examples like pyrochlore. However, Mielke did not seem to fully understand the degeneracy of his ground states, although he did know the singly doped ground states are degenerate. Recently, pyrochlore and checkerboard as special cases have been independently rediscovered by the authors of Ref.~\cite{pyrochlore}, who were unaware of Mielke's general results. They highlighted the spin liquid behavior of spin-charge separation in the singly hole-doped ground states, and for the first time fully characterized the ground state degeneracy for systems sizes of 16 and 32 sites, which will be explained within Mielke's graph theoretic framework in Appendix \ref{sec:Mielke}.

The existence of frustration-free ground states following this paradigm relies on the full symmetry among vertices of the simplices, which is only there under periodic boundary conditions. It also depends on the parity of the number of simplices along each axis of the lattice. It is expected that unshared corners on the boundary would result in ground state degeneracy as well as frustration.\footnote{Here frustration means that the ground state is not the common lowest energy eigenstate of each local Hamiltonian operator. It should be distinguished from the magnetic frustration referred to earlier in `kinetic frustration'.} Although each of the degenerate ground states can be chosen as a uniform superposition of all (monomer-)dimer tiling configurations, because the number of dimer coverings is not the same for different holon (and electron) locations, the holon distribution in the pyrochlore ground state is non-uniform.

The symmetry among the vertices is, however, not important for the singly doped sawtooth system and Husimi cacti, which also show RVB ground states where the holon instead hops in a valence bond solid (VBS) background~\cite{Katsura2015,PhysRevB.107.L140401}. This contrasting result was established by the `diamagnetic inequality' \cite{LiebFlux}, which makes it energetically favorable for the electrons in each triangle to form a spin singlet. Even though the dimer configuration is completely fixed by the holon location, the ground state is still not a uniform superposition of dimer configurations. But it is unique due to the Perron-Frobenius theorem since the effective Hamiltonian is a non-positive matrix. The approach relies on the fact that triangles are the only cycles in the lattice. For instance, the argument does not work for the kagome lattice, which contains additionally loops of length six among longer ones. In contrast, the longer cycles in the checkerboard and pyrochlore lattices with periodic boundary conditions do not destroy the RVB ground states.

The goal of the current paper is to generalize both results about RVB ground states on corner-sharing simplices, by studying the single-hole doped system on a chain of tetrahedra or a 1D stripe of the pyrochlore lattice. It can be viewed as an interpolation of the previous results, where neither are all corners of a tetrahedron shared with a neighboring tetrahedron nor do the spin-doublet and quadruplet sectors of a hole-doped tetrahedron share the same Hamiltonian that facilitates the proof of the RVB ground state using the diamagnetic inequality. In Sec.~\ref{sec:sawtooth}, we briefly review the counter-Nagaoka effect and its consequence on the sawtooth lattice. In addition, we solve the ground state energy and wavefunction analytically in the thermodynamic limit. Sec.~\ref{sec:pyro} generalizes the method used in the sawtooth model to study the pyrochlore stripe, which has localized monomer-dimer ground states. Sec.~\ref{sec:conclusion} concludes the article by pointing out a few follow-up directions.

\section{The RVB Ground State on the Sawtooth Lattice}
\label{sec:sawtooth}

\begin{figure}
	\centering
	\includegraphics[width=0.6\linewidth]{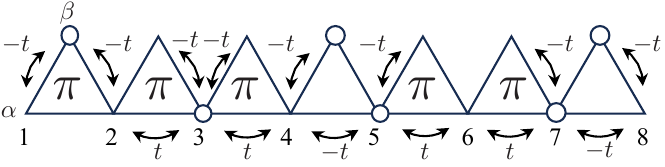}
	\caption{Effective hopping strength in the presence of vacancies. Due to the fermionic statistics, particle-hole transformation introduces a $\pi$-flux through odd-length loops containing a holon.}
	\label{fig:lattice}
\end{figure}

The quasi-1D sawtooth lattice can be viewed as a single stripe of the 2D kagome lattice, consisting of corner-sharing triangles. The unit cell consists of two sites: a lower vertex $\alpha$ of degree four (except at the boundary of the chain), and an upper vertex $\beta$ of degree two (see Fig.~\ref{fig:lattice}). The unit cells are labeled by an index $j=1,2,\ldots,L$. With open boundary conditions, there are $2L+1$ vertices (sites) in the lattice. On each lattice site, the local Hilbert space is spanned by three states: occupied by a $S=1/2$ fermion with spin up or down, or unoccupied. 

We study the Fermi-Hubbard model with infinite onsite potential $U$, with the low-energy effective Hamiltonian
\begin{equation}
H = -t\sum_{j=1}^L\sum_{\bm{\sigma}=\{\uparrow,\downarrow\}} P\left(c_{j\alpha,\sigma_{j\alpha}}^\dagger c_{j\beta,\sigma_{j\beta}}+c_{j\alpha,\sigma_{j\alpha}}^\dagger c_{(j+1)\alpha,\sigma_{(j+1)\alpha}} +c_{j\beta,\sigma_{j\beta}}^\dagger c_{(j+1)\alpha,\sigma_{(j+1)\alpha}}\right)P+\mathrm{h.c.},\label{eq:Ham}
\end{equation}where $t$ is the electron hopping amplitude and 
\begin{equation}
    P=\left(1-c^\dagger_{(L+1)\alpha,\uparrow}c^\dagger_{(L+1)\alpha,\downarrow}\right)\prod_{j=1}^L\prod_{\nu=\alpha,\beta}\left(1-c^\dagger_{j\nu,\uparrow}c^\dagger_{j\nu,\downarrow}\right)
\end{equation}is the Gutzwiller projection operator that forbids double occupancy. Due to fermionic statistics and the existence of odd-length loops in the lattice, the effective hopping strength depends on the number of vacancies in the loop. The presence of a single holon introduces a $\pi$-flux that can be chosen to reverse the sign of the hopping strength across the horizontal bond. An effective hopping strength is specified by labeling the lattice sites and choosing the Hilbert space basis
\begin{equation}
	|iv,\bm{\sigma}\rangle=c_{1\alpha,\sigma_{1\alpha}}^\dagger c_{1\beta,\sigma_{1\beta}}^\dagger c_{2\alpha,\sigma_{2\alpha}}^\dagger\cdots c_{(i-1)\beta,\sigma_{(i-1)\beta}}^\dagger c_{iv,\sigma_{iv}}^\dagger c_{(i+1)\alpha,\sigma_{(i+1)\alpha}}^\dagger\cdots c_{(L+1)\alpha,\sigma_{(L+1)\alpha}}^\dagger|\chi\rangle,
\end{equation}for $i=1,2,\dots, L, v=\alpha,\beta$ or $i=L+1,v=\alpha$. Here and in the remaining of the article, $|\chi\rangle$ denotes the pseudovacuum state with no electrons, and we label the lattice sites in ascending order from left to right, leading to the hopping constant for different holon configurations summarized in Fig.~\ref{fig:lattice}.

\begin{figure}
	\centering
	\includegraphics[width=0.6\linewidth]{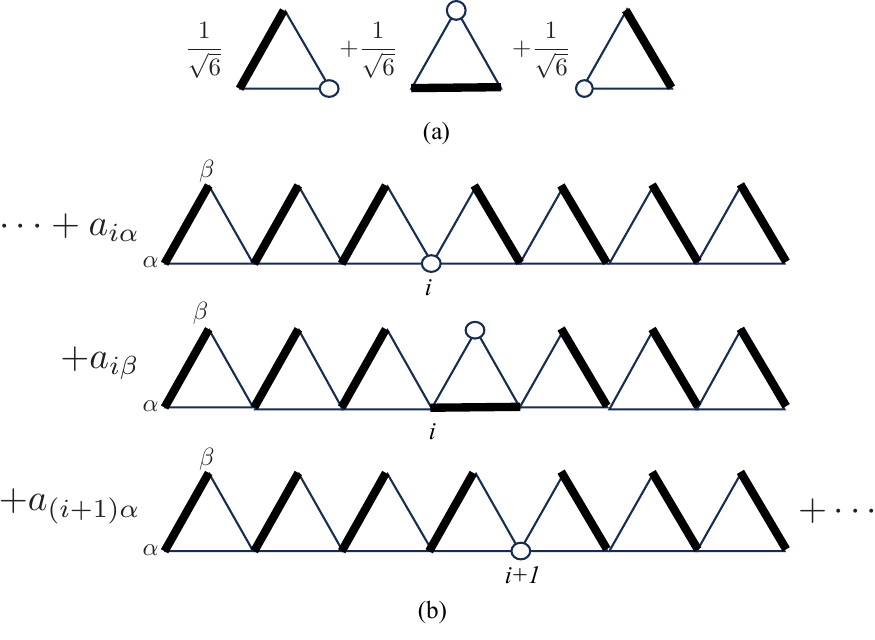}
	\caption{The ground states in the single holon sector of (a) a single triangle as a superposition of dimer configurations, and (b) the sawtooth lattice as a superposition of domain wall configurations.}
	\label{fig:GS}
\end{figure}

The single-hole doped many-body system can be approached by first understanding the few-body system of a single triangle, the ground state of which (for $t > 0$) is the equal superposition of the holon located at the three different corners with the spins at the remaining two corners forming a singlet (Fig.~\ref{fig:GS} (a)), with energy $E=-2t$.
%\footnote{The dimers have a direction as opposite directions are off by a minus sign in $|\uparrow_j\downarrow_k\rangle-|\downarrow_j\uparrow_k\rangle$. The convention in this subsection is to always have the dimers pointing in the clockwise direction.}
An effective hopping Hamiltonian for the holon on the triangle can be written down, which is block diagonal, separating the subspaces of the two spins forming a singlet and a triplet. The Hamiltonians in the two sectors turn out to be identical up to a minus sign. Next, the diamagnetic inequality \cite{LiebFlux} is used to show that the lowest energy eigenstate resides in the sector where the dimers in all the triangles of the sawtooth chain form spin singlets. Finally, the Perron-Frobenius theorem can be invoked to conclude that the ground state is a unique superposition of all dimer configurations \cite{PhysRevB.107.L140401}
\begin{equation}
	\begin{split}
		|\mathrm{GS}\rangle=& \sum_{i=2}^{L}\Big(a_{i\alpha}\prod_{j=1}^{i-1}d_{j\alpha,j\beta}^\dagger d_{i\beta,(i+1)\alpha}^\dagger\prod_{k=i+1}^{L}d_{k\beta,(k+1)\alpha}^\dagger+a_{i\beta}\prod_{j=1}^{i-1}d_{j\alpha,j\beta}^\dagger d_{(i+1)\alpha,i\alpha}^\dagger \prod_{k=i+1}^{L}d_{k\beta,(k+1)\alpha}^\dagger\Big) |\chi\rangle\\ 
		&+\Big(a_{1\alpha}d_{1\beta,2\alpha}^\dagger+a_{1\beta}d_{2\alpha,1\alpha}^\dagger\Big)\prod_{k=2}^{L}d_{k\beta,(k+1)\alpha}^\dagger |\mathrm{vac}\rangle+a_{(L+1)\alpha}\prod_{j=1}^{L}d_{j\alpha,j\beta}^\dagger|\chi\rangle,
	\end{split}	
\end{equation}where $d_{u,v}^\dagger=c_{u,\uparrow}^\dagger c_{v,\downarrow}^\dagger-c_{u,\downarrow}^\dagger c_{v,\uparrow}^\dagger$ represents a dimer singlet between neighboring vertices $u$ and $v$, and $a_{iv}$ denotes the probability amplitude of the dimer configuration with the holon located at the vertex $v$ of the $i$th unit cell. The holon marks a domain wall between two segments with dimers aligned in different orientations, as shown in Fig.~\ref{fig:GS} (b). We note that in this ground state, the hole position fixes a single dimer configuration. This RVB state is thus induced by a hole moving in a VBS.
\begin{figure}
	\centering
	\includegraphics[width=0.7\linewidth]{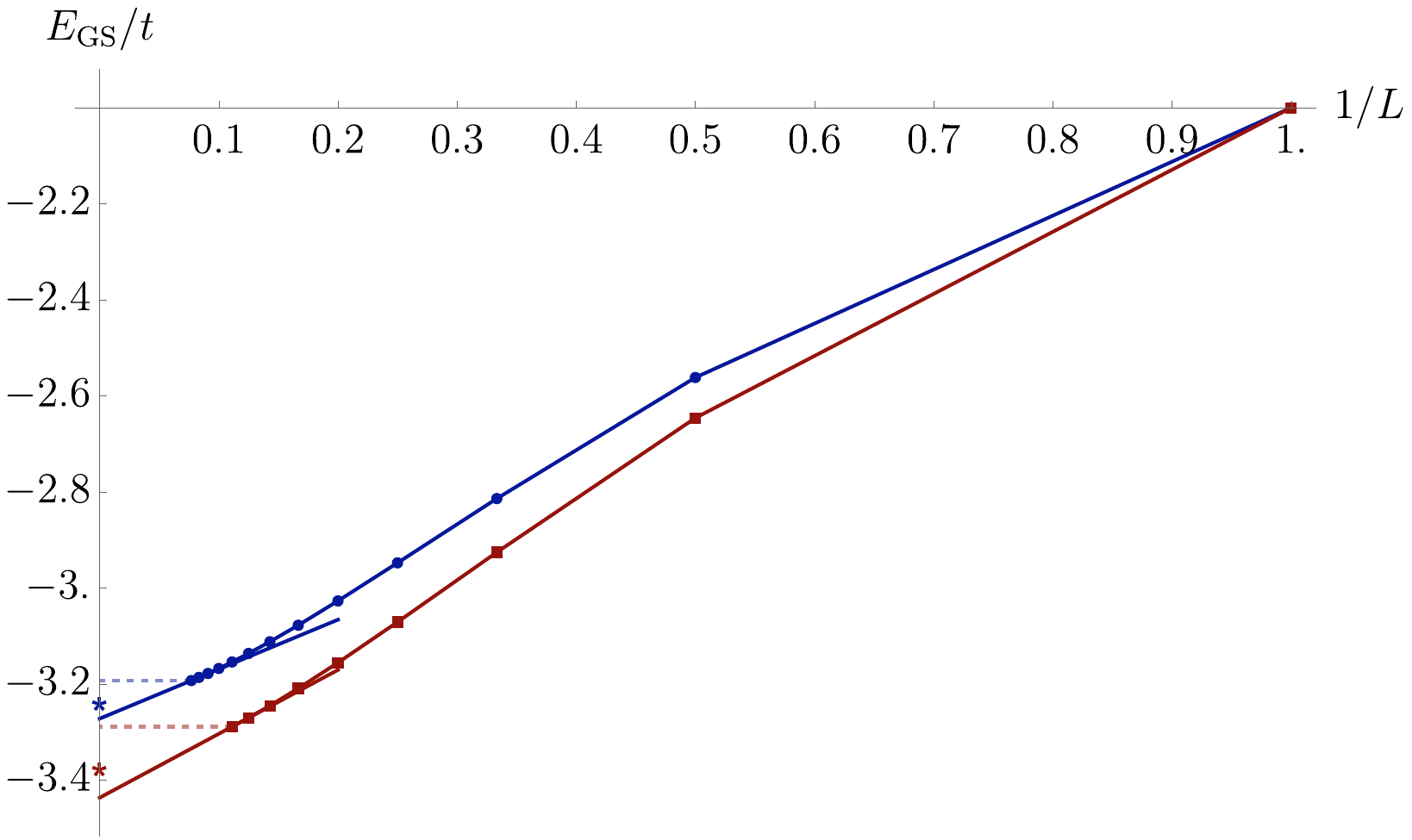}
	\caption{Ground state energy $E_{\rm GS}$ versus $1/L$ from exact diagonalization for the sawtooth chain (blue) and the pyrochlore stripe (red). Dashed lines give the ground state energy of the largest system studied, while solid lines give an extrapolation of $L\to\infty$ from the two biggest system sizes studied. Stars mark the analytical results for an infinite chain, which fall in between these two.}
	\label{fig:EGS_numerics}
\end{figure}

The coefficients $a_{i\alpha}, a_{i\beta}$ and the ground state energy can be determined analytically in the thermodynamic limit from the effective tight-binding Hamiltonian of the holon hopping in a background of singlet dimers
\begin{equation}
    H_\mathrm{eff}=-t\sum_j (b^\dagger_{j\alpha}b_{j\beta}+b^\dagger_{j\alpha}b_{(j+1)\alpha}+b^\dagger_{(j+1)\alpha}b_{j\beta})+\mathrm{h.c.},\label{eq:Heff}
\end{equation} where $b^\dagger_{j \nu}$ is the hole creation operator on sublattice $\nu$ in unit cell $j$. Using the translational invariance of the infinitely long chain, this effective model can be solved using the Fourier transform
\begin{equation}
    b_{j\nu}=\frac{1}{\sqrt{L}}\int_{-\pi}^\pi d\theta e^{ij\theta}b_{\theta\nu}.
\end{equation}In momentum space, the Hamiltonian \cref{eq:Heff} decouples into separate momentum sectors
\begin{equation}
    H_\mathrm{eff}=-t\int_{-\pi}^\pi d\theta \left(2\cos\theta b_{\theta\alpha}^\dagger b_{\theta \alpha}+(1+e^{-i\theta})b_{\theta\alpha}^\dagger b_{\theta\beta}+(1+e^{i\theta})b_{\theta \beta}^\dagger b_{\theta \alpha} \right).
\end{equation}
This gives the spectrum $E_\mathrm{eff}(\theta)=-t\big(\cos\theta\pm\sqrt{(\cos\theta+1)^2+1}\big)$, which is minimized at $\theta = 0$, making the ground state energy $E_{\rm GS}=-(\sqrt{5}+1)t\approx-3.23607t$. The ground state wavefunction in the thermodynamic limit is given by Fig.~\ref{fig:GS} with $a_{i\alpha}=\frac{\sqrt{5}+1}{2}$, $i=1,2,\dots, L+1$, and $a_{i\beta}=1$ for $i=1,2,\dots, L$. Its energy fits well with numerical simulations (exact diagonalization of the original Hamiltonian Eq.~\eqref{eq:Ham}), shown in Fig.~\ref{fig:EGS_numerics}.

\section{The Monomer-dimer Ground States on the Pyrochlore Stripe}
\label{sec:pyro}

\begin{figure}
	\centering
	\includegraphics[width=0.8\linewidth]{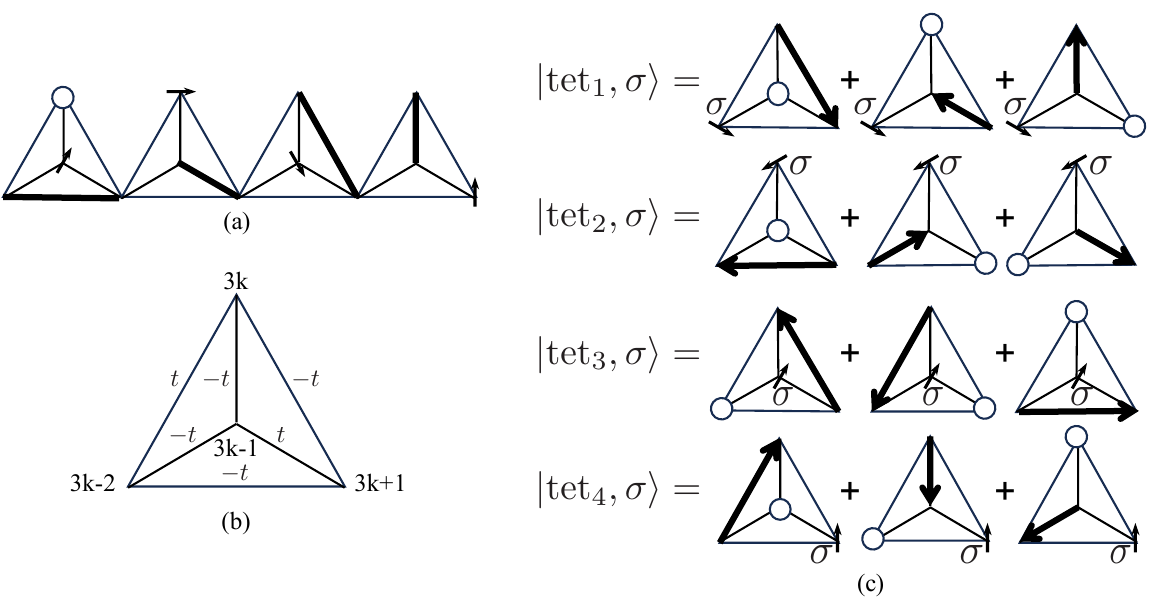}
	\caption{(a) A pyrochlore stripe consisting of 4 tetrahedra, showing one possible dimer configuration in the ground states having a single vacancy. (b) The effective hopping constants inside the $k$th tetrahedron, where the hopping constant between two sites $j$ and $j^\prime$ is defined by the convention $(-1)^{j-j^\prime}t$. (c) The four degenerate ground states of a single tetrahedron. The large arrows represent a spin-singlet dimer and the small arrows represent a spin-doublet. Unlike in the sawtooth lattice which allows a fixed convention of the dimer direction, the direction of the dimers here are specified by arrows.}
	\label{fig:stripe}
\end{figure}

Like the sawtooth lattice, the pyrochlore stripe or tetrahedron chain has two types of vertices, one of degree 6, the other of degree 3, see Fig.~\ref{fig:stripe} (a). It has been studied previously under the Heisenberg interaction in Ref.~\cite{PhysRevB.74.174424}, where under strong rung bonds, the ground state is shown to be a product of dimer states on the rung bonds and Bethe ansatz ground state of the Heisenberg chain. Although the model is different from the one of interest here, that result is also a consequence of the two different types of vertices. In this subsection, we examine how the RVB ground state of the single-hole doped 3D pyrochlore lattice~\cite{pyrochlore} gets modified when not all corners are shared between two tetrahedra.

An effective Hamiltonian of the infinite-$U$ Hubbard model on a single tetrahedron with a single vacancy is determined by specifying an ordering of the vertices in the Fock basis, as shown in Fig.~\ref{fig:stripe} (b). This Hamiltonian has been diagonalized for $t>0$ in Ref.~\cite{pyrochlore}, giving a ground state energy of $E=-2t$. The six-fold degenerate ground states are presented in Fig.~\ref{fig:stripe} (c), along with their linear combination $|\mathrm{tet}_4\rangle=-\sum_{i=1}^3|\mathrm{tet}_i\rangle$.\footnote{The dimers have a direction as opposite directions are off by a minus sign in $|\uparrow_j\downarrow_k\rangle-|\downarrow_j\uparrow_k\rangle$. Note that this definition of a dimer has the ordering of the sites explicitly accounted for, as opposed to the previous definition of $d_{j,k}^{\dagger}$.} Notice that they have not been orthonormalized. The four ground states are labeled by the location of the spin-doublet. In the triangular face opposite to the doublet, the holon and the dimer form the same ground state as the single triangle in the previous subsection. Note that $|\mathrm{tet}_{2,3}\rangle$ is different from $|\mathrm{tet}_{1,4}\rangle$ in that the triangle has two corners on the lattice sites of the 1D chain, instead of one. %Heuristically speaking, this results in degenerate ground states in the tetrahedron chain.

\subsection{Symmetries and Irreducible Representations}

\begin{figure}
	\centering
	\includegraphics[width=0.8\linewidth]{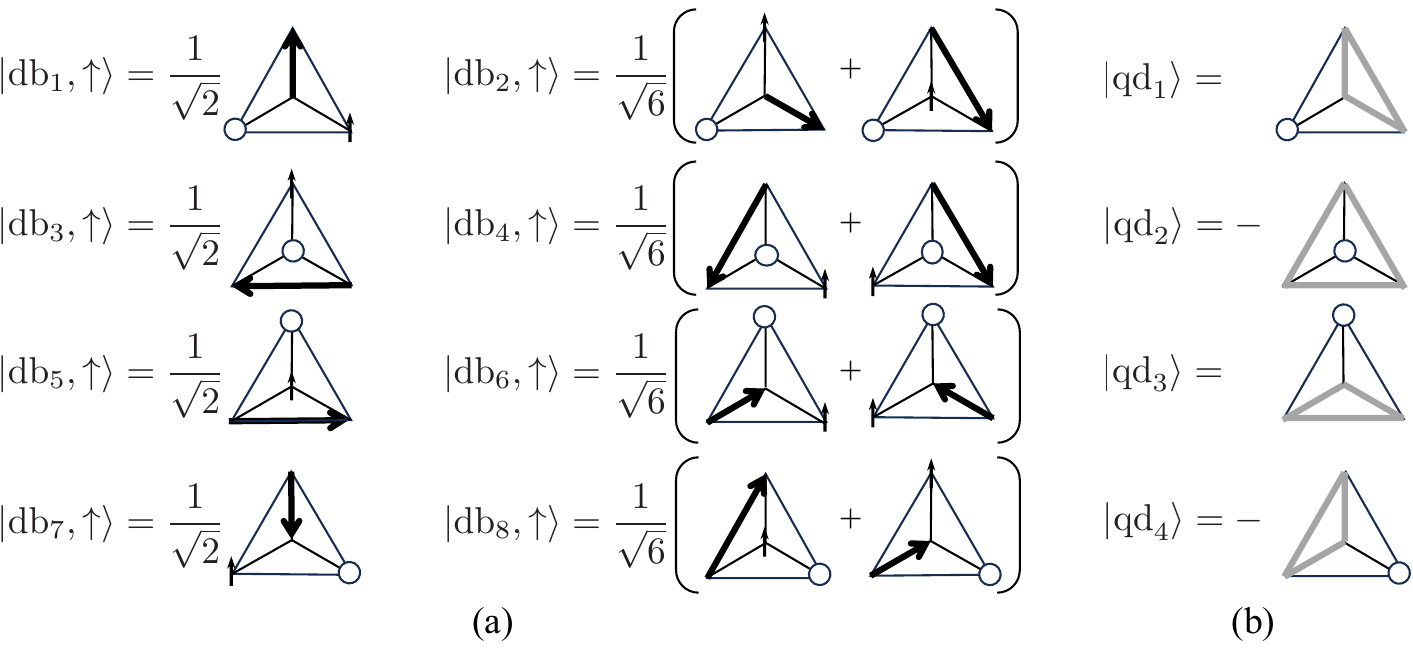}
	\caption{An orthonormal basis for the subspace with two spin-ups and one spin-down in (a) the 8-dimensional spin-doublet sector, and (b) the 4-dimensional spin-quadruplet sector. The grey triangle represents that the spins at the three corners are symmetrized.}
	\label{fig:doublet}
\end{figure}

The total spin operator $S$ and its $z$-component $S^z$ both commute with the hopping Hamiltonian. So, their eigenvalues are good quantum numbers. Furthermore, although a tetrahedron can consist of either three or four spins, there is an unambiguous way to group the spins into groups of three: For a given holon location, there are three vertices hosting spins within the one or two tetrahedra that contains the holon. After removing those already grouped spins, their neighboring tetrahedra also have only three spins left.\footnote{Note that this does not work for periodic boundary conditions, which will be reserved for future work.} As a result, the total spin of each group is also a conserved charge. The tensor product of three spin-$\frac{1}{2}$ representations of SU(2) is decomposed into a spin-$\frac{3}{2}$ quadruplet and two spin-$\frac{1}{2}$ doublets
\begin{equation*}
	\frac{1}{2}\otimes\frac{1}{2}\otimes\frac{1}{2}=\big(1\oplus 0\big)\otimes\frac{1}{2}= \frac{3}{2}\oplus\frac{1}{2}\oplus\frac{1}{2}.
\end{equation*}
For a fixed total $z$-component of the three spins $S^z_\mathrm{tet}=1/2$, the local Hilbert space of the three spins is hence three-dimensional, consisting of two dimer-monomer states listed in Fig.~\ref{fig:doublet} (a) ($S_\mathrm{tet}=1/2$), and one symmetrized state shown in Fig.~\ref{fig:doublet} (b) ($S_\mathrm{tet}=3/2$). The commutativity $[S_\mathrm{tet},H]=0$ says that the Hamiltonian will not mix subspaces with different $S_\mathrm{tet}$ for each tetrahedron. So $H$ is block diagonal in the irreducible representation specified by fixed values of $\{S_\mathrm{tet}\}$, where each $S_\mathrm{tet}=1/2$ or $3/2$:
\begin{equation}
    H=\sum_{k=1}^{L} H_k=\sum_{k=1}^{L} H_k^\mathrm{db}\oplus H_k^\mathrm{qd},\label{eq:tetraHam}
\end{equation}where 
\begin{equation}
    H_k^\mathrm{db(qd)}=\underbrace{1 \otimes 1 \otimes \cdots \otimes 1}_{3(k-1)}\otimes H^\mathrm{db(qd)} \otimes \underbrace{1 \otimes 1 \otimes \cdots \otimes 1}_{3(L-k)}.
\end{equation} The irreducible representations $H^\mathrm{db(qd)}$ for the spin-doublet (spin-quadruplet) sector $S_\mathrm{tet}=1/2$ ($S_\mathrm{tet}=3/2$) are given (in the bases defined in Fig.~\ref{fig:doublet}) by
\begin{equation}
	H^{\mathrm{db}}=t\begin{pmatrix}
		0_{2\times 2} & JKJ & KJ & I\\
		JKJ & 0_{2\times 2} & I & KJ \\
		JK & I & 0_{2\times 2} & K\\
		I & JK & K & 0_{2\times 2} 
		\end{pmatrix}, \quad H^{\mathrm{qd}}=t\begin{pmatrix}
		0 & 1& 1 & 1 \\ 1 & 0& 1 & 1\\ 1 & 1 & 0 & 1\\ 1 & 1 & 1 & 0 
	\end{pmatrix}, \label{eq:Hdb}	
\end{equation}where
\begin{equation}
	%R=\begin{pmatrix}-1/2& -\sqrt{3}/2\\ \sqrt{3}/2&  -1/2\end{pmatrix},\quad  
    I=\begin{pmatrix}1& 0\\ 0&  1\end{pmatrix},\quad J=\begin{pmatrix}-1& 0\\ 0&  1\end{pmatrix}, \quad K=\begin{pmatrix}1/2& -\sqrt{3}/2\\ -\sqrt{3}/2&  -1/2\end{pmatrix},
\end{equation}satisfy $J^2=K^2=(JK)^3=I$, forming a two-dimensional representation of the symmetric group $\mathcal{S}_3$. $H^{\mathrm{db}}$ has eigenvalues $\pm 2t$ (three-fold degenerate) and $0$ (doubly degenerate), whereas $H^{\mathrm{qd}}$ has eigenvalues $3t$ and $-t$ (three-fold degenerate).

\subsection{Energy Level Ordering of Irreducible Representations}

A state $|\{\psi_h\};\{S_k\};\{\bm{\phi}_k\}\rangle$ in the representation specified by $\{S_k\}$ is parametrized by the probability amplitudes $\{\psi_h\}$ to find the holon at site $h$, and angles $\{\bm{\phi}_k\}$ for the tetrahedra that are in the $S_k=1/2$ irreducible representation as
\begin{equation}
    |\{\psi_h\};\{S_k\};\{\bm{\phi}_k\}\rangle=\sum_{h=1}^{3L+1}\psi_h|h;\{S_k\};\{\bm{\phi}_k\}\rangle=\sum_{h=1}^{3L+1}\psi_h|0\rangle_h\bigotimes_{k=1}^L|S_k\rangle,
\end{equation}where
\begin{equation}   
   |S_k=\frac{1}{2}\rangle=\begin{cases}
    \cos\phi_k^{(1)}|\mathrm{db_1},\uparrow\rangle+\sin\phi_k^{(1)}|\mathrm{db_2},\uparrow\rangle, \\ \cos\phi_k^{(2)}|\mathrm{db_3},\uparrow\rangle+\sin\phi_k^{(2)}|\mathrm{db_4},\uparrow\rangle, \\\cos\phi_k^{(3)}|\mathrm{db_5},\uparrow\rangle+\sin\phi_k^{(3)}|\mathrm{db_6},\uparrow\rangle,\\ \cos\phi_k^{(4)}|\mathrm{db_7},\uparrow\rangle+\sin\phi_k^{(4)}|\mathrm{db_8},\uparrow\rangle,\\
    \end{cases} |S_k=\frac{3}{2}\rangle=\begin{cases}|\mathrm{qd}_1\rangle,& h=3k-2,\\ |\mathrm{qd}_2\rangle,&h=3k-1,\\ |\mathrm{qd}_3\rangle,& h=3k,\\ |\mathrm{qd}_4\rangle,&h=3k+1,\end{cases}
\end{equation}for the $k$th tetrahedron hosting the holon.

Its energy $E(\{\psi_h\};\{S_k\};\{\bm{\phi}_k\})=\langle\{\psi_h\};\{S_k\};\{\bm{\phi}_k\}|H|\{\psi_h\};\{S_k\};\{\bm{\phi}_k\}\rangle$ can be written as
\begin{equation}
\begin{split}
    E(\{\psi_h\};\{S_k\};\{\bm{\phi}_k\})=&\sum_{k=1}^{L} E_k(\bm{\psi}_k;S_k;\bm{\phi}_k)\\ =&\sum_{k=1}^{L}\sum_{h,h'=3k-2}^{3k+1}\psi^*_{h'}\psi_h\langle h';\{S_k\};\{\bm{\phi}_k\}|H_k|h;\{S_k\};\{\bm{\phi}_k\}\rangle.
\end{split}
\end{equation}
In light of the Lieb-Mattis theorem on energy ordering by the pouring principle \cite{PhysRev.125.164}, it will not be too surprising if it turns out that for any given probability amplitudes $\bm{\psi}_k$ of finding the holon at the four vertices of the $k$th tetrahedron, there exists a choice of $\bm{\phi}_k$ such that the energy contributions of the $k$th tetrahedron satisfy
\begin{equation}
    E_k(\bm{\psi}_k;\frac{1}{2};\bm{\phi}_k)<E_k(\bm{\psi}_k;\frac{3}{2}).\label{eq:inequality}
\end{equation}In the following, we will prove this using a non-Abelian generalization to the diamagnetic inequality \cite{LiebFlux} used in Ref.~\cite{PhysRevB.107.L140401} analogously for the sawtooth chain.

Since the Hamiltonian \cref{eq:tetraHam} is a real symmetric matrix, its eigenvectors can all be chosen to have real components, so we can assume $\psi_h$'s are real, for $1\le h \le 3L+1$. The energy of the spin-quadruplet sector (right hand side of \cref{eq:inequality}) is given by
\begin{equation}
\begin{split}
    E_k(\bm{\psi}_k;\frac{3}{2})=& 2t\left(p_k^{(1)}+\frac{p_{k}}{p_{k}^{(1)}}+p_k^{(2)}+\frac{p_{k}}{p_{k}^{(2)}}+p_k^{(3)}+\frac{p_{k}}{p_{k}^{(3)}} \right),
    %\\ =& 2t\sum_{k=1}^{L} \Big(p_{k}^{(1)}+\frac{p_{k}}{p_k^{(1)}}+p_{k}^{(2)}+\frac{p_{k}}{p_k^{(2)}}+p_{k}^{(3)}+\frac{p_{k}}{p_k^{(3)}}\Big),
\end{split}
		\label{eq:equad}
\end{equation}where $p_k=\psi_{3k-2}\psi_{3k-1}\psi_{3k}\psi_{3k+1}$, and $p_k^{(i)}=\psi_{3k-2}\psi_{3k-2+i}$, for $i=1,2,3$. On the other hand, the energy of the spin-doublet sector (left hand side of \cref{eq:inequality}) can be expressed as
\begin{equation}
	\begin{split}
		 E_k(\bm{\psi}_k;\frac{1}{2};\bm{\phi}_k)
         =2t\Bigg(&p_k^{(1)}\cos\varphi_{k}^{(1)}+\frac{p_{k}}{p_{k}^{(1)}}\cos\left(\varphi_{k}-\varphi_{k}^{(1)}+\frac{2\pi}{3}\right) \\ +&p_{k}^{(2)}\cos\varphi_{k}^{(2)}+\frac{p_{k}}{p_{k}^{(2)}}\cos\left(\varphi_{k}-\varphi_{k}^{(2)}-\frac{2\pi}{3}\right)\\ +&p_{k}^{(3)}\cos\varphi_{k}^{(3)}+\frac{p_{k}}{p_{k}^{(3)}}\cos\left(\varphi_{k}-\varphi_{k}^{(3)}\right)\Bigg), 
	\end{split}\label{eq:edoublet}
\end{equation}where
\begin{equation}
\begin{aligned}
    \varphi_k=&-\phi_k^{(1)}-\phi_k^{(2)}+\phi_k^{(3)}+\phi_k^{(4)}, \\
    \varphi_k^{(1)}=&-\phi_k^{(1)}-\phi_k^{(2)}+\frac{\pi}{3},   \\
    \varphi_k^{(2)}=&-\phi_k^{(1)}+\phi_k^{(3)}+\frac{2\pi}{3}, \\ 
    \varphi_k^{(3)}=&-\phi_k^{(1)}+\phi_k^{(4)}.
\end{aligned}\label{eq:angles}
\end{equation}
\begin{figure}
	\centering
	\includegraphics[width=0.6\linewidth]{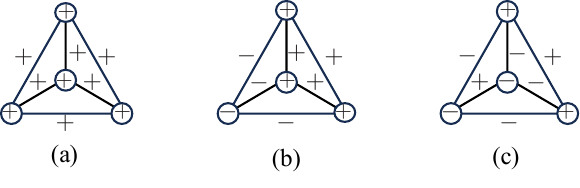}
	\caption{The three possible scenarios for different number of negative terms in Eq.~\eqref{eq:equad}. The vertices denote the sign of a single probability amplitude, whereas the edges represent the sign of the product of two such amplitudes.}
	\label{fig:sings}
\end{figure}

There are only three different scenarios for the signs of the six terms in \cref{eq:equad}, as summarized in Fig.~\ref{fig:sings}. In the scenario shown in Fig.~\ref{fig:sings} (a), all six terms are positive, so after shrinking by factors of cosines in \cref{eq:edoublet}, the inequality \cref{eq:inequality} is always guaranteed. It is a strict inequality because all of the variables of the cosines cannot simultaneously be integer multiples of $2\pi$, as the sums of the pair of the arguments of the cosines in each line of \cref{eq:edoublet} are respectively $\varphi_{k}+\frac{2\pi}{3}$, $\varphi_{k}-\frac{2\pi}{3}$, and $\varphi_{k}$. In the second scenario, Fig.~\ref{fig:sings} (b), we need to have three of the cosine factors in \cref{eq:edoublet} corresponding to the three negative terms in \cref{eq:equad} to be simultaneously -1. For instance, if the vertex with negative probability amplitude is $3k-2$, this can be done by setting $\varphi_k^{(i)}=0$ for $i=1,2,3$, which is possible since we have three independent degrees of freedom to tune for a given value of $\phi_k^{(1)}$. Additionally, $\phi_k^{(1)}$ can still be tuned to make the remaining three cosines smaller than 1. For the last case, Fig.~\ref{fig:sings} (c), without loss of generality, we can suppose the last four terms in \cref{eq:equad} are negative. In order to show that the there exists choices of the angles in \cref{eq:angles} that realize \cref{eq:inequality}, we first tune the three independent variables $\varphi_k^{(i)}, i=1,2,3$, to get the minimum value of \cref{eq:edoublet} as a function of $\varphi_k$ only
\begin{equation}
\begin{split}
    E_k(\bm{\psi}_k;\frac{1}{2};\varphi_k)=&\min_{-\phi_k^{(1)}-\phi_k^{(2)}+\phi_k^{(3)}+\phi_k^{(4)}=\varphi_k} E_k(\bm{\psi}_k;\frac{1}{2};\bm{\phi}_k) \\ =&-2t\Bigg(\sqrt{a_k^2+2p_k\left(\cos(\varphi_k+\frac{2\pi}{3})-1\right)} +\sqrt{b_k^2+2p_k\left(\cos(\varphi_k-\frac{2\pi}{3})-1\right)}\\ &\qquad +\sqrt{c_k^2+2p_k\left(\cos(\varphi_k)-1\right)}\Bigg) ,
\end{split}
\end{equation}where (using the inequality of arithmetic and geometric means)
\begin{equation}
    a_k=p_k^{(1)}+\frac{p_{k}}{p_{k}^{(1)}}\ge 2\sqrt{p_k}, \quad b_k=p_k^{(2)}+\frac{p_{k}}{p_{k}^{(2)}}\le -2\sqrt{p_k}, \quad c_k=p_k^{(3)}+\frac{p_{k}}{p_{k}^{(3)}}<0.
\end{equation}
Then we can let $\varphi_k=0$, so that we only need to show $\sqrt{a_k^2-3p_k} +\sqrt{b_k^2-3p_k}>|a_k+b_k|$. Squaring both sides, this is equivalent to $\sqrt{(a_k^2-3p_k)(b_k^2-3p_k)}> 3p_k+a_kb_k$, which trivially holds since the RHS is negative.

So the conclusion is that in the ground state, the tri-spin group in each tetrahedron in the chain must form spin-doublets, meaning that the ground states are superpositions of configurations with $L$ dimers, one holon, and $L$ spin-doublet monomers. However, unlike the sawtooth chain and Husimi cactus studied in Ref.~\cite{PhysRevB.107.L140401}, because the matrix elements of $H^\mathrm{db}$ are not all negative, we cannot deduce that the ground state is unique in the $S_\mathrm{tot}^z=L/2$ subspace. We will dig more into this degeneracy in the next subsection.

\begin{figure}
	\centering
	\includegraphics[width=0.4\linewidth]{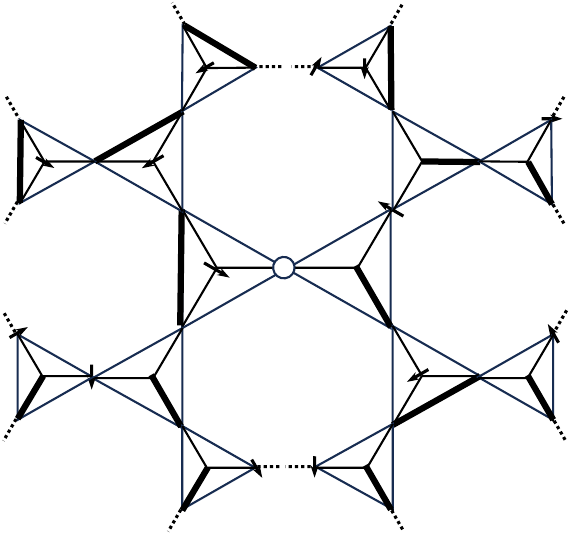}
	\caption{A partial RVB configuration on the Bethe lattice of tetrahedra with coordination number $z=3$.}
	\label{fig:tree}
\end{figure}

In fact, we have also shown that the groups of three spins inside each tetrahedron always form SU(2)-doublets in the ground state of the single hole-doped Hubbard model at $U=\infty$ on any network of corner-sharing tetrahedra, each of which having 1 to 4 neighboring tetrahedra, as long as there are no cycles involving more than one tetrahedron. This is necessary for a well defined division of the vertices into groups of three starting from the location of the vacancy. Among these networks are the regular trees or Bethe lattices of coordination numbers $z=2,3,4$, with $z=2$ corresponding to the pyrochlore stripe. The $z=3$ lattice is depicted in Fig.~\ref{fig:tree}, where the nodes of the tree are tetrahedra, and the edges are the shared corners.

The conclusion of this subsection is verified with numerical exact diagonalization results for pyrochlore stripes of size up to $L=7$, by computing the expectation value of the total spin of the three spins residing in the triangular planes excluding the holon. Like for the sawtooth case, where once the location of the vacancy is fixed, the dimers are all aligned in parallel lines with different orientation on each side of the holon, here the doublets occupy the triangular faces in the same way. However, a difference is that for a given hole position, there are multiple possible such dimer coverings, i.e. the hole moves in an RVB background.

\subsection{Exact Ground States in the Thermodynamic Limit}\label{sec:pyroED}

Like in the case of the sawtooth chain, the ground state can be solved analytically using translational invariance in the thermodynamic limit. But unlike the sawtooth chain, within each total $S^z$ sector, there is a two-dimensional local Hilbert space for a holon residing at vertices $3k-1$ or $3k$, and a four-dimensional subspace for a holon located at vertices $3k-2$ and $3k+1$. So two adjacent tetrahedra are connected by a unitary transformation $\bm{v_{3k+1}}=\mathcal{U}\bm{v_{3k-2}}$ between the basis vectors $\bm{v_4}=v_4^{(1)}|\mathrm{db}_7,\sigma\rangle_k +v_4^{(2)}|\mathrm{db}_8,\sigma\rangle_k$ and $\bm{v_1}=v_1^{(1)}|\mathrm{db}_1,\sigma\rangle_{k+1} +v_1^{(2)}|\mathrm{db}_2,\sigma\rangle_{k+1}$, such that the probability $|\bm{v_1}|^2=|\bm{v_4}|^2$ of finding the holon at vertices along the 1D chain is well-defined. It turns out that the lowest energy eigenstates are found in the subspace respecting the symmetry $v_1^{(1)}=\pm v_4^{(1)}, v_1^{(2)}=\pm v_4^{(2)}$ between vertices 1 and 4, corresponding to $\mathcal{U}=\pm I$ or $\pm \sigma_z$.  Each of them defines a distinct translationally invariant subspace containing a lowest energy eigenstate, leading to a four-fold degenerate ground state for a fixed total $S^z$.

Denoting the components of the eigenstate in the bases of $|\mathrm{db}_{1,2},\uparrow\rangle$, $|\mathrm{db}_{3,4},\uparrow\rangle$, and $|\mathrm{db}_{5,6},\uparrow\rangle$ in Fig.~\ref{fig:doublet} by $\bm{v_{1,2,3}}$ respectively, the translationally invariant ground state in the thermodynamic limit can be solved from the eigenvalue equation 
\begin{equation}
    \begin{pmatrix}
        \ddots & \vdots & \vdots & \vdots & \vdots & \vdots & \vdots & \vdots & \cdots\\
        \cdots & JKJ & 0_{2\times 2} & I & KJ\mathcal{U} & \multicolumn{3}{c}{\multirow{2}{*}{
        \scalebox{2}{$0_{4\times 6}$}}} & \cdots\\
        \cdots & JK & I &  0_{2\times 2} & K\mathcal{U} & \multicolumn{3}{c}{} & \cdots\\
        \cdots & \mathcal{U} & \mathcal{U}JK & \mathcal{U}K &  0_{2\times 2} & JKJ & KJ & \mathcal{U} & \cdots\\
        \cdots & \multicolumn{3}{c}{\multirow{2}{*}{\scalebox{2}{$0_{4\times 6}$}}} & JKJ &  0_{2\times 2} & I & KJ\mathcal{U} & \cdots\\
        \cdots & \multicolumn{3}{c}{} & JK & I &  0_{2\times 2} & K\mathcal{U} & \cdots\\
        \cdots & \vdots & \vdots & \vdots & \vdots & \vdots & \vdots & \vdots & \ddots\\
    \end{pmatrix} \begin{pmatrix}\vdots\\ e^{-i\theta}\bm{v_1} \\ e^{-i\theta}\bm{v_2} \\ e^{-i\theta}\bm{v_3} \\ \bm{v_1} \\ \bm{v_2} \\ \bm{v_3} \\ e^{i\theta}\bm{v_1} \\ \vdots \end{pmatrix} =\frac{E}{t}\begin{pmatrix}\vdots\\ e^{-i\theta}\bm{v_1} \\ e^{-i\theta}\bm{v_2} \\ e^{-i\theta}\bm{v_3} \\ \bm{v_1} \\ \bm{v_2} \\ \bm{v_3} \\ e^{i\theta}\bm{v_1} \\ \vdots \end{pmatrix},
\end{equation}
which has six independent equations
\begin{equation}
    \left\{ \begin{aligned}
		   \left(2\cos\theta \mathcal{U}- \frac{E}{t} \right)\bm{v_1}+\left(e^{-i\theta}\mathcal{U}JK+JKJ\right)\bm{v_2} +\left(e^{-i\theta}\mathcal{U}K +KJ\right)\bm{v_3}  =&\begin{pmatrix}0\\ 0\end{pmatrix},\\
			\left(JKJ+e^{i\theta}KJ\mathcal{U} \right)\bm{v_1}-\frac{E}{t}\bm{v_2 }+\bm{v_3} =&\begin{pmatrix}0\\ 0\end{pmatrix},\\ \left(JK+e^{i\theta}K\mathcal{U} \right)\bm{v_1}+\bm{v_2 }-\frac{E}{t}\bm{v_3} =&\begin{pmatrix}0\\ 0\end{pmatrix}.
		\end{aligned}
		\right.
\end{equation}
For $\mathcal{U}=I$, the energy eigenvalues are determined by the vanishing of the determinant of the coefficient matrix 
\begin{equation}\renewcommand\arraystretch{1.5}
    \left|\begin{matrix} 2\cos\theta-\frac{E}{t} & 0 & \frac{1}{2}(1-e^{-i\theta}) &  \frac{\sqrt{3}}{2}(1+e^{-i\theta}) &  -\frac{1}{2}(1-e^{-i\theta}) &  -\frac{\sqrt{3}}{2}(1+e^{-i\theta})\\ 
    0 & 2\cos\theta-\frac{E}{t} &  \frac{\sqrt{3}}{2}(1-e^{-i\theta}) & -\frac{1}{2}(1+e^{-i\theta})  &  \frac{\sqrt{3}}{2}(1-e^{-i\theta}) & -\frac{1}{2}(1+e^{-i\theta}) \\ 
    \frac{1}{2}(1-e^{i\theta}) & \frac{\sqrt{3}}{2}(1-e^{i\theta}) & -\frac{E}{t} & 0 &1 & 0 \\ 
   \frac{\sqrt{3}}{2}(1+e^{i\theta})  &-\frac{1}{2}(1+e^{i\theta})  &  0 & -\frac{E}{t} & 0& 1\\    
   -\frac{1}{2}(1-e^{i\theta})  & \frac{\sqrt{3}}{2}(1-e^{i\theta}) & 1 & 0& -\frac{E}{t} & 0  \\ 
   -\frac{\sqrt{3}}{2}(1+e^{i\theta}) &-\frac{1}{2}(1+e^{i\theta})  &0& 1&  0 & -\frac{E}{t}
    \end{matrix} \right|.
\end{equation}The determinant vanishes when $E=\pm t$, as the last four rows and columns are linearly dependent for all values of the momentum $\theta$. These flat bands are due to the effective decoupling of vertices $3k-1$ and $3k$ from the rest of the chain when they are (anti-)symmetrized by requiring $\bm{v_2}=\pm J\bm{v_3}$. Exploiting this symmetry to simplify the determinant, we find the other four bands
\begin{equation}
\frac{E}{t}=\cos\theta +\frac{1}{2}\pm\sqrt{(\cos\theta-\frac{3}{2})^2+2}, \cos\theta -\frac{1}{2}\pm\sqrt{(\cos\theta+\frac{3}{2})^2+2}.
\end{equation} The energy is minimized at $\theta=\pi$, giving the ground state energy $E_{\rm GS} \approx -3.37228t$. We show in Fig..~\ref{fig:EGS_numerics} that this result is consistent with extrapolation of the ground state energy from finite sized systems using numerical exact diagonalization.
%The result for $L=9$ gives the upper bound $E_{\rm GS}^{\rm max} = -3.28854t$, while extrapolation from $L=8$ and $L=9$ gives a lower bound of $E_{\rm GS}^{\rm min} = -3.43633t$.

\begin{figure}
	\centering
	\includegraphics[width=0.5\linewidth]{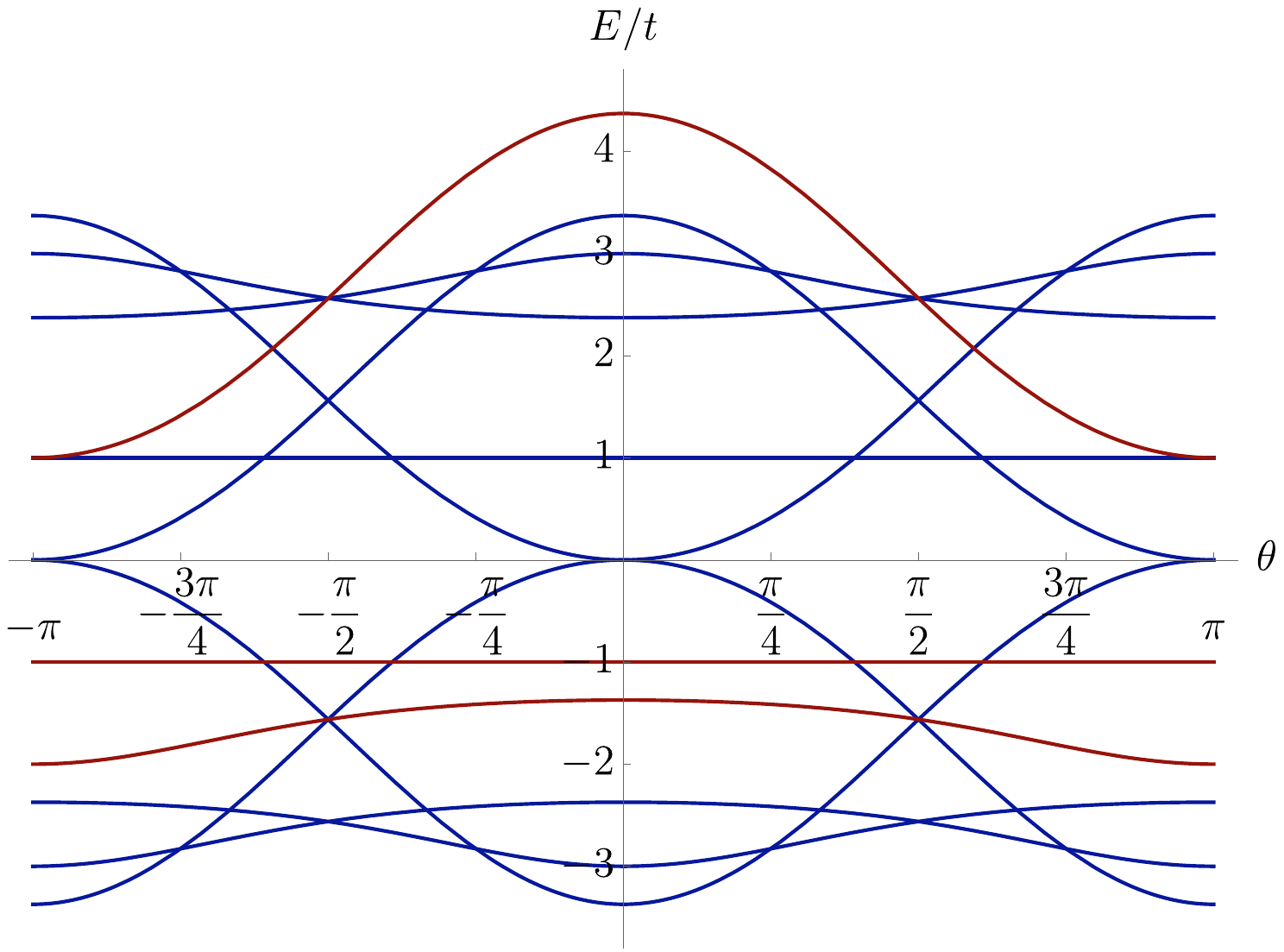}
	\caption{Band diagram of the infinite-$U$ Hubbard model with a single vacancy on an infinitely long pyrochlore stripe in the four irreducible subspaces containing the degenerate ground states corresponding to $U=\pm I, \pm \sigma_z$, where all tetrahedra form spin-doublets (blue). The subspace containing the highest energy eigenstate, where all tetrahedra form spin-quadruplets (red) are plotted for comparison. The flat band at $E=-t$ appears in both sectors.}
	\label{fig:bd}
\end{figure}

The energy bands in the other three sectors are listed below
\begin{equation}
    \frac{E}{t}=\begin{cases}\pm1,
        -\cos\theta +\frac{1}{2}\pm\sqrt{(\cos\theta+\frac{3}{2})^2+2}, -\cos\theta -\frac{1}{2}\pm\sqrt{(\cos\theta-\frac{3}{2})^2+2}, & \mathcal{U}= -I; \\ \pm1,
        -\cos\theta +\frac{1}{2}\pm\sqrt{(\cos\theta+\frac{3}{2})^2+2}, \cos\theta -\frac{1}{2}\pm\sqrt{(\cos\theta+\frac{3}{2})^2+2}, & \mathcal{U}= \sigma_z, \\ \pm1,
        \cos\theta +\frac{1}{2}\pm\sqrt{(\cos\theta-\frac{3}{2})^2+2}, -\cos\theta -\frac{1}{2}\pm\sqrt{(\cos\theta-\frac{3}{2})^2+2}, & \mathcal{U}= -\sigma_z. \\ 
    \end{cases}
\end{equation}
The 10-band spectrum in the infinite chain limit is plotted in Fig.~\ref{fig:bd}.
For comparison, we also show the three-band spectrum of the quadruplet sector in the same figure.
There, the energies are given by
\begin{equation}
\frac{E}{t}=- 1, \cos\theta +\frac{1}{2}\pm\sqrt{(\cos\theta+\frac{3}{2})^2+2}.
\end{equation}

In the infinite-$U$ limit, ground states are present in all total $S^z$ sectors where the three spins within each tetrahedron form a spin-doublet. This is because the Hamiltonian commutes with the total $S^z$ operator defined for each group of three spins inside a tetrahedron. That accounts for a ground state degeneracy of $\binom{L}{L/2+S^z_\mathrm{tot}}$ for the subspace with total $S^z_\mathrm{tot}$ in the entire chain, which sums to $2^{L}$ for the $L+1$ different possible values of $S^z_\mathrm{tot}$. 

Large, but finite, $U$ allows for an effective AFM Heisenberg interaction in addition to the hopping ($t$-$J$ model). This interaction does not commute with the total $S^z$ operator within each group of three spins for individual tetrahedra. Moreover, unlike the sawtooth case, its lowest energy on a single tetrahedron is achieved when the four spins form two singlet dimers. Thus for a chain of tetrahedra, the ground state of the AFM Heisenberg interaction is frustrated and competing with the hopping Hamiltonian. The ground states reside in the sector with minimal total spin, either a total spin-singlet or doublet depending on the parity of $L$, corresponding to a four- or eight-fold degeneracy, respectively.

\section{Conclusion}\label{sec:conclusion}

In this article, the RVB ground states of the singly hole-doped Hubbard model on the sawtooth and pyrochlore lattices are generalized to resonating mixtures of dimers and monomers on a quasi-1D chain of tetrahedra. The ground state degeneracy scales exponentially with the lattice size in the limit of infinite onsite potential $U$, and is lifted to a finite four- or eight-fold degeneracy at finite $U$ by the Heisenberg interaction in the effective $t$-$J$ model. Since the argument that leads to the partial RVB nature of the ground states is independent on the lattice geometry except the requirement that there are no cycles that involve multiple tetrahedra, the result also applies to any such networks of corner-sharing tetrahedra with different numbers of unshared corners.

Our preliminary result seems to point to a more general paradigm of understanding many-body ground states on corner-sharing lattices by studying the energy level ordering of a few-body system, in our case a simplex structure like triangle or tetrahedron. The solution of the many-body problem in the latter example bears resemblance to non-Abelian gauge theory or general relativity as basis vectors of the global Hilbert space are constructed by an atlas of charts that map the local bases at the shared corner of neighboring tetrahedra to each other. It is worth investigating in the future whether one can find sufficient or necessary conditions on the local Hamiltonian that is block diagonalized into different spin irreducible representations for the global ground state to have a particular representation of total spin.   

In order to further push our result to two- and three-dimensional corner-sharing lattices, the first step is to understand the effect of periodic boundary conditions on the quasi-1D lattices. It is easy to see that in this case the holon can hop from either side to the same simplex, necessarily delocalizing the dimers, which is closer to Anderson's original proposal of the RVB state. The current argument of the RVB nature of the ground state would not apply any more, although numerical evidence indicates that much of our result still holds. A similar problem appears when more holons are introduced to the quasi-1D lattices. Although certain models with two holons can be solved with the coordinate Bethe ansatz~\cite{Gayen_1995}, next-nearest neighbor hopping is generally believed break integrability, which is necessary for the diagonalization of systems with three or more holons.

However, even if the finite doping problem could somehow be attacked using nested Bethe ansatz, the solution would not be manifestly capturing the dimer structure. It is therefore interesting to see if there exists an interpolating method that combines the advantage of the real space and momentum space analysis. One related construction is the partially integrable states of $S_N$ symmetric chains, which is solved by applying the coordinate Bethe ansatz to a background of pre-antisymmetrized basis of $S_N$ singlets \cite{PhysRevB.106.134420}. Another successful example is the application of Bethe ansatz on top of a matrix product state (MPS) and the generalized frustration-free model \cite{PhysRevB.109.104307}. The nice thing about dimerized ground states is that they can usually be described by MPS or its higher dimensional counterparts, from which a transfer matrix can be obtained to compute the correlation function \cite{correlation}. However, for our boundary condition, the MPS would not be translationally invariant, which makes it less useful for analytic calculations.

Last but not least, it is interesting to see whether there are possible generalizations on lattices composed of higher simplex structures beyond triangles and tetrahedra. One can also consider more local degrees of freedom. In Ref.~\cite{PhysRevResearch.6.013307}, such generalizations are made when the number of degrees of freedom on a lattice site grows at the rate of the unit cell of the lattice. It would be more physically relevant if the result can be generalized to lattices with larger unit cells while the local Hilbert space dimension remains 3, or at least grows at a slower rate. Along these lines, the results of diagonalizing the $S_k$ version of the XX model on $k$-simplices might become relevant \cite{ZHANG2023169395}.

\section*{Acknowledgements}
We thank Olav F. Syljuåsen for discussions. ZZ thanks Hosho Katsura and Jeffrey Teo for discussions.

% TODO: include author contributions
\paragraph{Author contributions}
ZZ designed the project, conducted the theoretical part of the research and drafted the manuscript. CG performed the numerical work and contributed to writing.

\paragraph{Funding information}
CG acknowledges funding from the European Union’s Horizon Europe research and innovation programme under the Marie Skłodowska-Curie grant agreement No. 101126636.

\begin{appendix}
\numberwithin{equation}{section}

\section{The Brandt-Giesekus Frustration-free Ground States} \label{sec:BG}

In this appendix, we review the exactly solvable ground states of the infinite-$U$ Hubbard model on corner-sharing simplices with periodic boundary conditions. They are frustration free after rewriting the Hamiltonian using the operator identities \cite{PhysRevLett.68.2648}
\begin{equation}
    \begin{split}
        -Pc^\dagger_{r',\sigma}c_{r,\sigma}P=&c_{r,\sigma}Pc^\dagger_{r',\sigma},\quad \mathrm{for}\ r'\ne r,\\
       P(1-n_{r,\sigma}-n_{r,-\sigma})P=& c_{r,\sigma}Pc^\dagger_{r,\sigma},
    \end{split}\label{eq:identites}
\end{equation}where the number operator $n_{r,\sigma}=c^\dagger_{r,\sigma}c_{r,\sigma}$ satisfies the commutation relations
\begin{equation}
    [n_{r,\sigma},c^\dagger_{r',\sigma'}]=\delta_{r,r'}\delta_{\sigma,\sigma'}c^\dagger_{r',\sigma'},\quad [n_{r,\sigma},c_{r',\sigma'}]=-\delta_{r,r'}\delta_{\sigma,\sigma'}c_{r',\sigma'}.
\end{equation}The Gutzwiller projection operator $P=\prod_r(1-n_{r,\uparrow}n_{r,\downarrow})$ then satisfy the commutation relation
\begin{equation}
    [c^\dagger_{r,\sigma},P]=c^\dagger_{r,\sigma}n_{r,-\sigma}P, \quad [P,c_{r,\sigma}]=Pn_{r,-\sigma}c_{r,\sigma}.
\end{equation}
Using these relations, we have for $r'\ne r$,
\begin{equation}
\begin{split}
        -Pc^\dagger_{r',\sigma}c_{r,\sigma}P=& Pc_{r,\sigma}c^\dagger_{r',\sigma}P\\
    =&(c_{r,\sigma}P+Pn_{r,-\sigma}c_{r,\sigma})(Pc^\dagger_{r',\sigma}+c^\dagger_{r',\sigma}n_{r',-\sigma}P)\\
    =&c_{r,\sigma}Pc^\dagger_{r',\sigma}+Pn_{r,-\sigma}c_{r,\sigma}Pc^\dagger_{r',\sigma}+c_{r,\sigma}Pc^\dagger_{r',\sigma}n_{r',-\sigma}P+Pn_{r,-\sigma}c_{r,\sigma}c^\dagger_{r',\sigma}n_{r',-\sigma}P.
\end{split}
\end{equation}
Since $n_{r,-\sigma}c_{r,\sigma}P=Pc^\dagger_{r',\sigma}n_{r',-\sigma}=0$, the last three terms vanish identically. Otherwise, using the property of projection operators $P^2=P$, the RHS becomes
\begin{equation}
    \begin{split}
        c_{r,\sigma}PPc^\dagger_{r,\sigma}=&Pc_{r,\sigma}(1-n_{r,-\sigma})(1-n_{r,-\sigma})c^\dagger_{r,\sigma}P\\
        =&Pc_{r,\sigma}(1-n_{r,-\sigma})c^\dagger_{r,\sigma}P\\
        =&P(1-n_{r,\sigma})(1-n_{r,-\sigma})P\\
        =&P(1-n_{r,\sigma}-n_{r,-\sigma})P,
    \end{split}
\end{equation}where in the last step the identity $Pn_{r,\sigma}n_{r,-\sigma}P=0$ has been used.

Brandt and Giesekus originally considered $D$-dimensional decorated cubic lattices, but their reformation of the Hamiltonian readily genearlizes to other lattices of corner-sharing $k$-simplices without boundaries, such as the pyrochlore lattice ($k=3$). The lattices studied by Brandt and Giesekus are thus special cases for $k=2D-1$. Using \eqref{eq:identites}, the Hamiltonian can be written in terms of the emergent degree of freedom within the $R$th simplex $S_R$
\begin{equation}
    \psi^\dagger_{R,\sigma}=\frac{1}{\sqrt{k+1}}\sum_{r\in S_R}c^\dagger_{r,\sigma} \label{eq:cell}
\end{equation}as
\begin{equation}
    \begin{split}
        H=&-t\sum_{r\ne r'}\sum_{\sigma}Pc^\dagger_{r',\sigma}c_{r,\sigma}P\\
        =&t\sum_{r\ne r'}\sum_{\sigma}c_{r,\sigma}Pc^\dagger_{r',\sigma}\\
        =&(k+1)t\sum_{R}\sum_{\sigma}\psi_{R,\sigma}P\psi^\dagger_{R,\sigma}-2t\sum_{r}\sum_{\sigma}c_{r,\sigma}Pc^\dagger_{r,\sigma}\\
        =&(k+1)t\sum_{R}\sum_{\sigma}\psi_{R,\sigma}P\psi^\dagger_{R,\sigma}-2t\sum_{r}\sum_{\sigma}P(1-n_{r,\sigma}-n_{r,-\sigma})P\\
        =&(k+1)t\sum_{R}\sum_{\sigma}\psi_{R,\sigma}P\psi^\dagger_{R,\sigma}-4tP(N_r-N_c)P,
    \end{split}\label{eq:BGHam}
\end{equation}where $N_r$ is the number of lattice sites, and $N_c=\sum_{r,\sigma}n_{r,\sigma}$ is the total number of electrons.
For fixed particle number, the second term of \eqref{eq:BGHam} is a constant, while the first term is positive semi-definite. So the ground state energy is lower-bounded by $E_\mathrm{GS}\ge -4tN_h$, with $N_h$ being the number of holons. This lower bound is achieved by the state
\begin{equation}
    |\mathrm{BG}\rangle =P\prod_{R}\psi^\dagger_{R,\uparrow}\psi^\dagger_{R,\downarrow}|\chi\rangle.
\end{equation}This state is annihilated by the first term in \eqref{eq:BGHam} due to the operator identity $P\psi^\dagger_{R,\sigma}P\psi^\dagger_{R,\sigma}=0$.

Tasaki observed that $|\mathrm{BG}\rangle$ is actually an RVB state \cite{PhysRevLett.70.3303} by rewriting it as
\begin{equation}
    |\mathrm{BG}\rangle=\frac{1}{k+1}P\prod_R\sum_{r< r'\in S_R}(c^\dagger_{r,\uparrow}c^\dagger_{r',\downarrow}-c^\dagger_{r,\downarrow}c^\dagger_{r',\uparrow})|\chi\rangle.
\end{equation}This is a superposition of dimer covering configurations with one dimer within each simplex. For corner-sharing $k$-simplex lattices without boundaries, each simplex contains $(k+1)/2$ lattice sites. On the other hand, at least two electrons per simplex cell are needed so as to form a dimer. So Brandt and Giesekus concluded that that non-trivial results only happen for hypercubic lattices with dimension $D\ge 3$, which for corner-sharing $k$-simplices means $k\ge 4$, as the checkerboard lattice ($D=2, k=3$) and the pyrochlore lattice ($k=3$) would be half-filled. However, this restriction has been significantly alleviate by Mielke later on \cite{Mielke_1992}.

Before we review Mielke's generalization in Appendix.~\ref{sec:Mielke}, let us first discuss what happens when not all corners of each simplex is shared with a neighbor. For such lattices with boundaries, the second term in the Hamiltonian \eqref{eq:BGHam} has to be modified from a constant to
\begin{equation}
    -t\sum_{\sigma}\left(2\sum_{r\in\Lambda^\circ }c_{r,\sigma}Pc^\dagger_{r,\sigma}+\sum_{r\in \partial\Lambda}c_{r,\sigma}Pc^\dagger_{r,\sigma}\right)=-tP(2N_\circ+N_\partial-2n_\circ-n_\partial)P,
\end{equation}where $\Lambda^\circ$ (resp.~$\partial\Lambda$) denotes the bulk (resp.~boundary) of the lattice, and $N_\circ$ (resp.~$N_\partial$) are the total number of lattice sites in the bulk shared between two simplices (resp.~on the boundary), with $n_{\circ}, n_{\partial}$ being the corresponding electron number operators. This term favors more electrons residing on the boundary corners, which competes with the positive semidefinite first term in \eqref{eq:BGHam} in the presence of doped holons. So the ground state is no longer frustration free, as we have seen in the main text.\footnote{In Ref.~\cite{PhysRevB.52.2476}, Giesekus also considered the tetrahedron chain, but with an additional onsite potential for the unshared vertices on the boundary, so that the Hamiltonian remains of the form \eqref{eq:BGHam}, resulting in the frustration-free ground state with an energy that is an integer multiple of $t$.}

\section{Mielke's Graph Theoretic Generalization}\label{sec:Mielke}

Mielke reformulated corner-sharing simplex lattices as the line graphs (also known as the adjoint or conjugate lattices) of the lattices of centers of the simplices. For example, the checkerboard lattice is the line graph of the square lattice, and the pyrochlore lattice is the line graph of the diamond lattice. The vertices of the line graph $L(G)$ of a graph $G$ are the edges of $G$, and the vertices are adjacent if their corresponding edges in $G$ share an endpoint. In this language, $r$ and $R$ in the Hamiltonian \eqref{eq:BGHam} label respectively the vertices of $L(G)$ and $G$. Mielke's insight is that the linear dependence of the dimer basis on $L(G)$ can be studied with the \textit{incidence matrix} of $G$. 

For an undirected graph $G$ with $V$ vertices and $E$ edges, the incidence matrix $B$ is an $V\times E$ matrix with element 
\begin{equation}
    B_{ij}=\begin{cases}1, & \text{if the $i$th vertex is incident with the $j$th edge;}\\
        0, & \text{otherwise}.
    \end{cases}
\end{equation}Because each column of $B$ contains only two $1$s, the row vectors $\bm{b}_i$ are linearly dependent by $\sum_i \alpha_i \bm{b}_i=0$, for $\alpha_k=-\alpha_l$ if the $k$th and $l$th vertices are connected by an edge. This means that for connected graphs, we have $\mathrm{rank}(B)=V-1$. The rank-nullity theorem then tells us that the kernel of $B$ is $(E-V+1)$-dimensional. It is easy to see that the kernel of $B$ is spanned by the cycles of $G$, so this is nothing but Euler's formula.

Brandt and Giesekus' rewriting of the Hamiltonian \eqref{eq:BGHam} used the fact that the adjacency matrix of $L(G)$ can be expressed as 
\begin{equation}
    \begin{split}
        A(L(G))=B(G)^T B(G)-2 I_{E}=\sum_{R=1}^V \bm{b}^T_R \bm{b}_R-2I_E
    \end{split}
\end{equation}where $I_{E}$ is the identity matrix of dimension $E$. It becomes evident that the operators $P\psi^\dagger_{R,\sigma}$ in \eqref{eq:BGHam}, in two-to-one mapping with the row vectors $ \bm{b}_R$, are also linearly dependent for bipartite $G$ due to $\sum_R (-1)^R \psi^\dagger_{R,\sigma}=0$. So the big surprise for Brandt and Giesekus is that it suffices for the ground state to be annihilated by all these Hamiltonian terms except for one $R$. That is why the ground states for singly and doubly hole-doped systems are also frustration free. Mielke went on to prove that the ground state for two vacancies
\begin{equation}
    |R\rangle = P\prod_{R'\ne R} \psi^\dagger_{R',\uparrow}\psi^\dagger_{R',\downarrow}|\chi\rangle, \label{eq:2hGS}
\end{equation}for any $R$ is nonvanishing for bipartite $G$. We shall not repeat his proof here. Interested readers are instead directed to Ref.~\cite{Mielke_1992}.

Mielke was not able to show the uniqueness of this ground state. But it should be obvious that the states $|R\rangle$ are identical regardless of the choice of $R$, since for $R_1\ne R_2$ we have
\begin{equation}
    \begin{split}
        |R_2\rangle =& P\psi^\dagger_{R_1,\uparrow}\psi^\dagger_{R_1,\downarrow}\prod_{R\ne R_{1,2}}\psi^\dagger_{R,\uparrow}\psi^\dagger_{R,\downarrow}|\chi\rangle\\
        =& P\sum_{R'\ne R_1}(-1)^{R'-R_1}\psi^\dagger_{R',\uparrow}\sum_{R''\ne R_1}(-1)^{R''-R_1}\psi^\dagger_{R'',\downarrow}\prod_{R\ne R_{1,2}}\psi^\dagger_{R,\uparrow}\psi^\dagger_{R,\downarrow}|\chi\rangle\\
        =&P\psi^\dagger_{R_2,\uparrow}\psi^\dagger_{R_2,\downarrow}\prod_{R\ne R_{1,2}}\psi^\dagger_{R,\uparrow}\psi^\dagger_{R,\downarrow}|\chi\rangle\\
        =&|R_1\rangle.
    \end{split}
\end{equation}Nevertheless, he did know that the ground states with one holon
\begin{equation}
    |r,\sigma\rangle = Pc^\dagger_{r,\sigma}\prod_{R'\ne R} \psi^\dagger_{R',\uparrow}\psi^\dagger_{R',\downarrow}|\chi\rangle\label{eq:1hGS}
\end{equation}is degenerate. Like \eqref{eq:2hGS}, they are annihilated by the first term in \eqref{eq:BGHam} because\footnote{It should be noted that the analytical proof in the Supplementary Material A.~2 and 3 is incomplete. Without utilizing the topological property of the lattice, the correctly identified ground state has not been shown to saturate the lower bound on ground state energy, even in the thermodynamic limit.}
\begin{equation}
\begin{split}
    \sum_{R}\sum_{\sigma}\psi_{R,\sigma}P\psi^\dagger_{R,\sigma}P\prod_{R'\ne R_1} \psi^\dagger_{R',\uparrow}\psi^\dagger_{R',\downarrow}=&\sum_{\sigma}\psi_{R,\sigma}P\psi^\dagger_{R,\sigma}P\prod_{R'\ne R} \psi^\dagger_{R',\uparrow}\psi^\dagger_{R',\downarrow}\\
    =&-\sum_{\sigma}\psi_{R,\sigma}\sum_{R_1\ne R} (-1)^{R_1-R}P\psi^\dagger_{R_1,\sigma}P\prod_{R'\ne R} \psi^\dagger_{R',\uparrow}\psi^\dagger_{R',\downarrow}\\
    =&0.
\end{split}
\end{equation}In fact, its degeneracy can be counted from the nullity of the incidence matrix $B$. As we have already shown the invariance of $|r,\sigma\rangle$ on the choice of $R$ on the RHS of \eqref{eq:1hGS}, there are at most as many such ground states as the number of lattice sites $N_r$. But because of $\sum_{r\in R} c^\dagger_{r,\sigma}=\sqrt{k+1}\psi^\dagger_{R,\sigma}$, there are only $N_r-N_R+1$ linearly independent ones, where the extra 1 comes from the double counting due to the PBC. This is exactly the nullity of $B$ for the lattice of simplex centers. For instance, a pyrochlore lattice with $E$ sites has $V=E/2$ tetrahedra for periodic boundary condition, so the kernel of the positive semidefinite part of the Hamiltonian \eqref{eq:BGHam} is $E/2+1$ dimensional. Taking into account the double degeneracy due to the spin-doublet, the ground state for the singly hole-doped system is $E+2$. This has been numerically confirmed by one of the authors \cite{pyrochlore}.

\end{appendix}

\bibliography{pyro.bib}
\end{document}